\documentclass[a4paper,11pt]{article}
\usepackage{pos}

\usepackage{caption}
\usepackage{subcaption}
\usepackage{physics}

\newcommand{\RNum}[1]{\uppercase\expandafter{\romannumeral#1}}

\title{Lattice field theory results for hybrid static potentials at short quark-antiquark separations and their parametrization}
\ShortTitle{Hybrid static potentials at short quark-antiquark separations}

\author*[a,b]{Carolin Schlosser}
\author[a]{Sonja Köhler}
\author[a,b]{Marc Wagner}

\affiliation[a]{Institut für Theoretische Physik, Goethe-Universität Frankfurt am Main,\\
  Max-von-Laue-Straße 1, D-60438
  Frankfurt am Main, Germany}

\affiliation[b]{Helmholtz Research Academy Hesse for FAIR, Campus Riedberg, \\
	 Max-von-Laue-Straße 12, D-60438 
	 Frankfurt am Main, Germany}

\emailAdd{schlosser@itp.uni-frankfurt.de}
\emailAdd{skoehler@itp.uni-frankfurt.de}
\emailAdd{mwagner@itp.uni-frankfurt.de}

\abstract{
	We present SU(3) lattice Yang-Mills data for hybrid static potentials from five ensembles with different small lattice spacings and the corresponding parametrizations for quark-antiquark separations $0.08\,\text{fm} \le r \le 1.12\,\text{fm}$. We remove lattice discretization errors at tree level of perturbation theory and partly at order $a^2$ as well as the $a$-dependent self energy. In particular the tree-level improvement of static potentials is discussed in detail and two methods are compared. The resulting parametrizations are expected to represent continuum limit results for hybrid static potentials within statistical errors.
	}

\FullConference{%
	The 39th International Symposium on Lattice Field Theory,\\
	8th-13th August, 2022,\\
	Rheinische Friedrich-Wilhelms-Universität Bonn, Bonn, Germany
}

\begin{document}
\maketitle


	\section{Introduction}
	The goal of our work is to compute precise results for the hybrid static potentials $\Pi_u$ and $\Sigma_u^-$ at short quark-antiquark separations using SU(3) lattice Yang-Mills theory.
	The hybrid static potentials $\Pi_u$ and $\Sigma_u^-$ describe the two lowest excitations of the gluon field surrounding a static quark-antiquark pair as a function of the quark-antiquark separation.
	For the related heavy hybrid mesons, exotic quantum number combinations  $J^{PC}$ are possible due to the non-trivial quantum numbers of the gluonic excitation. 
	They are -- like tetraquarks and and glueballs -- an active field of research in experiments as well as in theory (for recent reviews see e.g.\ \cite{Olsen:2017bmm,Braaten:2014ita,Meyer:2015eta,Swanson:2015wgq,Lebed:2016hpi,Brambilla:2019esw}).
	Lattice field theory results for hybrid static potentials are an essential input for mass calculations of heavy $\bar b b$ or $\bar c c$ hybrid mesons within the Born-Oppenheimer approximation~\cite{Perantonis:1990dy,Juge:1999ie,Guo:2008yz,Braaten:2014qka,Capitani:2018rox}.
	Refined Born-Oppenheimer approaches also include heavy quark spin effects or the coupling of different channels~\cite{Berwein:2015vca,Oncala:2017hop,Brambilla:2019jfi}.

	For reliable predictions of heavy hybrid meson masses within such approaches precise lattice field theory results and parametrizations for the corresponding static potentials are important.
	To improve on existing lattice field theory computations of hybrid static potentials~ \cite{Perantonis1989StaticPF,Michael:1990az,Perantonis:1990dy,Juge:1997nc,Juge:1997ir,Capitani:2018rox,Juge:2002br,Bali:2003jq,Juge:2003ge},
	we use lattice spacings significantly smaller than those used in existing works.
	This allows us to remove lattice discretization errors at tree level of perturbation theory and to some extent at leading order in $a^2$.
	We obtain improved lattice results and parametrizations consistent with continuum limit results for the hybrid static potentials $\Pi_u$ and $\Sigma_u^-$ within statistical errors.

	
	\section{Lattice setup}
	We generated five ensembles of pure SU(3) gauge field configurations with the CL2QCD software package~\cite{Philipsen:2014mra} using a Monte Carlo heatbath algorithm and the standard Wilson plaquette action.
	\begin{table}[b]
		\begin{center}
			\def\arraystretch{1.1}
			\begin{tabular}{cccc}
				\hline
				ensemble & $\beta$ & $a$ in $\text{fm}$ \cite{Necco:2001xg} & $(L/a)^3 \times T/a$ \\
				\hline
				$A$ & $6.000$ & $0.093$ & $12^3\times 26$  \\
				$B$ & $6.284$ & $0.060$ & $20^3\times 40$ \\
				$C$ & $6.451$ & $0.048$ & $26^3\times 50$ \\ 
				$D$ & $6.594$ & $0.040$ & $30^3\times 60$  \\ \hline 
				$A^{\text{HYP}2}$ & $6.000$ & $0.093$  & $24^3 \times 48$\\
				\hline
			\end{tabular}
		\end{center}
		\caption{Gauge link ensembles.}
		\label{tab:latticesetups4}
	\end{table}
	The ensembles $A,\, B,\, C$ and $D$ (listed in Table~\ref{tab:latticesetups4}) were generated with gauge couplings \\ $\beta=6.000,\,6.284,\,6.451$ and $6.594$, which correspond to lattice spacings \\ $a=0.093\,\text{fm},\, 0.060\,\text{fm},\, 0.048\,\text{fm}$ and $0.040\,\text{fm}$, respectively. 
	The scale is set according to a parametrization of $\ln(a/r_0)$ from Ref.\ \cite{Necco:2001xg} and by identifying $r_0$ with $0.5\,\text{fm}$.
	To achieve a reduction of statistical errors, a multilevel algorithm~\cite{Luscher:2001up} was used on ensembles $A,\, B,\, C$ and $D$.
	We confirmed that finite volume effects are negligible for the physical lattice volumes of ensembles $A,\, B,\, C$ and $D$, which are $ T \times L^3 \approx 2.4\,\text{fm} \times (1.2\,\text{fm})^3$.
	Moreover, we checked that the measurements are neither affected by large autocorrelations nor topology freezing.
	For details on the exclusion of systematic errors from the finite volume, topology and also possible decays into glueballs see Ref.~\cite{Schlosser:2021wnr}.
	
	We also include lattice results from ensemble $A^\text{HYP2}$, which were obtained in the context of a preceeding project and publication~\cite{Capitani:2018rox}.
	The lattice spacing is the same as for ensemble $A$ but, due to the larger lattice volume, larger quark-antiquark separations up to $r \le 1.12\,\text{fm}$ were accessible.
	In contrast to results from ensembles $A,\, B,\, C$ and $D$, results from ensemble $A^\text{HYP2}$ were computed with HYP2-smeared temporal links~\cite{Hasenfratz:2001tw,DellaMorte:2003mn,DellaMorte:2005nwx}, which reduces the self energy of the static quarks but increases discretization errors at small $r/a$.


	\section{Lattice field theory computation of (hybrid) static potentials}
	We compute hybrid static potentials from Wilson loop-like correlation functions on the five ensembles $e \in \{A,\,B,\,C,\,D,\,A^\text{HYP2}\}$.
	In contrast to standard Wilson loops related to the ordinary static potential with quantum numbers $\Lambda_{\eta}^{\epsilon}=\Sigma_g^+$, the spatial link paths of the hybrid Wilson loops include suitable insertions to generate the non-trivial quantum numbers of hybrid static potentials.
	The quantum numbers $\Lambda_{\eta}^{\epsilon}$  denote the orbital angular momentum along the quark-antiquark separation axis, $\Lambda =  \Sigma (=0), \Pi (=1), \Delta (=2), \ldots$, the behavior under combined parity and charge conjugation, $\eta= g (=+), u (=-)$, and the behavior under reflection along an axis orthogonal to the quark separation axis, $\epsilon = +,-$.
	To compute the potentials $\Pi_u$ and $\Sigma_u^-$, we use the insertions $S_{\text{\RNum{3}},1}$ and $S_{\text{\RNum{4}},2}$, which were defined and optimized for maximal ground state overlaps in Ref.~\cite{Capitani:2018rox}.
	To further increase the ground state overlaps, spatial links are smeared with APE-smearing  on all ensembles with $\alpha_\text{APE}=0.5$, where the number of smearing steps $N_\text{APE}$ was increased with decreasing lattice spacing (see Ref.~\cite{Schlosser:2021wnr} for details).
	
\begin{figure}
		\begin{minipage}{0.6\linewidth}\raggedleft
			\centering\includegraphics[width=0.8\linewidth]{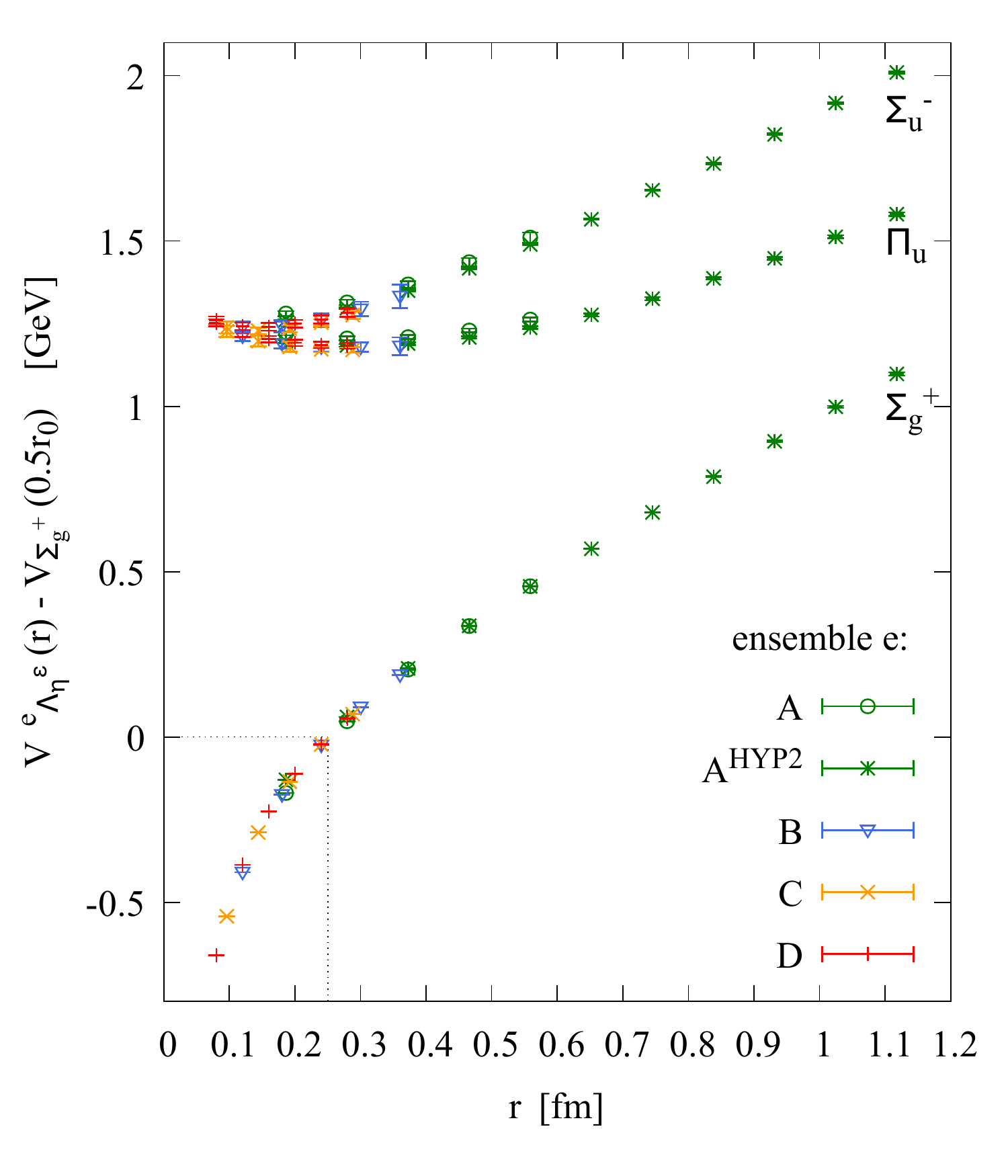}
			\caption{SU(3) lattice field theory results for the ordinary static potential $\Sigma_g^+$ and the hybrid static potentials $\Pi_u$ and $\Sigma_u^-$ from the five ensembles $e \in \{A,\,B,\,C,\,D,\,A^\text{HYP2}\}$.}
			\label{fig:unimproved_potentials}
		\end{minipage}
		\begin{minipage}{0.4\linewidth}\centering
			\centering\includegraphics[width=\linewidth]{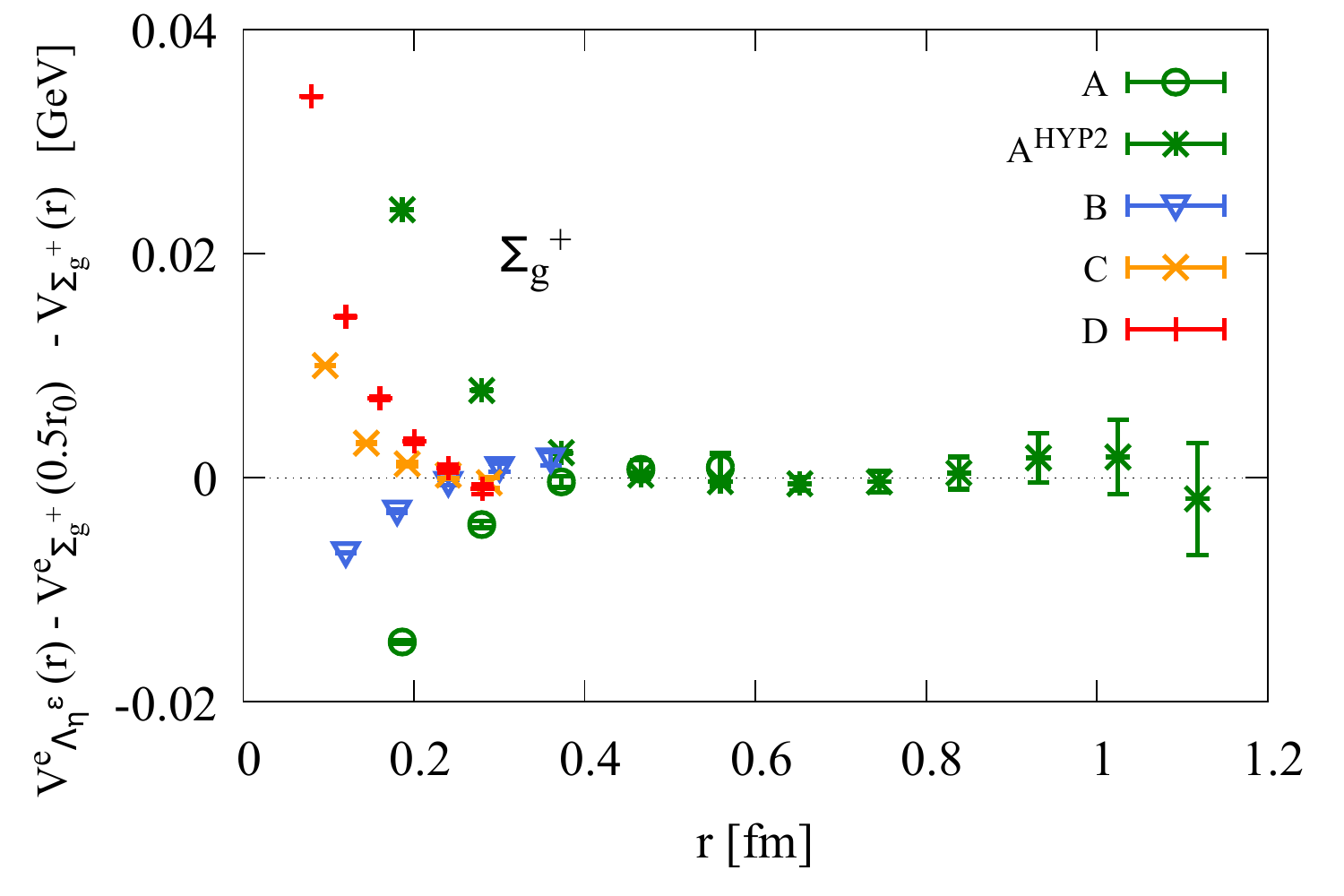}
			\caption{Visualization of lattice discretization errors in the lattice data for $\Sigma_g^+$. $V_{\Sigma_g^+}(r)$ as defined in Eq.~\eqref{eq:parametrization_Ordinary} is subtracted from the data shown in Figure~\ref{fig:unimproved_potentials}.}
			\label{fig:visualization_discr_err}
		\end{minipage}
	\end{figure}

	The static potentials $aV_{\Lambda_{\eta}^{\epsilon}}^e(r)$ are extracted by plateau fits of the corresponding effective potentials at large temporal separations of the Wilson loops. 
	Our SU(3) lattice Yang-Mills theory results for the ordinary static potential and the two lowest hybrid static potentials are presented in Figure~\ref{fig:unimproved_potentials}.
	To show the lattice data from all ensembles together in a meaningful plot in Figure~\ref{fig:unimproved_potentials}, we set $V^e_{\Sigma_g^+}(r=0.5r_0)=0$ for $e \in \{A,\,B,\,C,\,D,\,A^\text{HYP2}\}$ to compensate for the ensemble-dependent self energy of the static quarks.
	
	At our level of statistical precision, lattice discretization errors lead to large discrepancies between lattice results from different ensembles covering the same range of physical quark-antiquark separations. 
	This is visualized in Figure~\ref{fig:visualization_discr_err} for the ordinary static potential. 
	In the following section~\ref{SEC004} we discuss how to remove discretization errors to a large extent. 
	After that, in section~\ref{SEC005}, we are able to provide a common parametrization of all available lattice data sets representing their continuum limit.

	
	\section{\label{SEC004}Comparsion of two methods of tree-level improvement for the static potential}
\begin{figure}[htb] 
			\begin{center}
			\begin{minipage}{0.5\linewidth}
				\subcaption{no improvement}\label{fig:unimproved}
				\centering\includegraphics[width=0.8\linewidth,page=1]{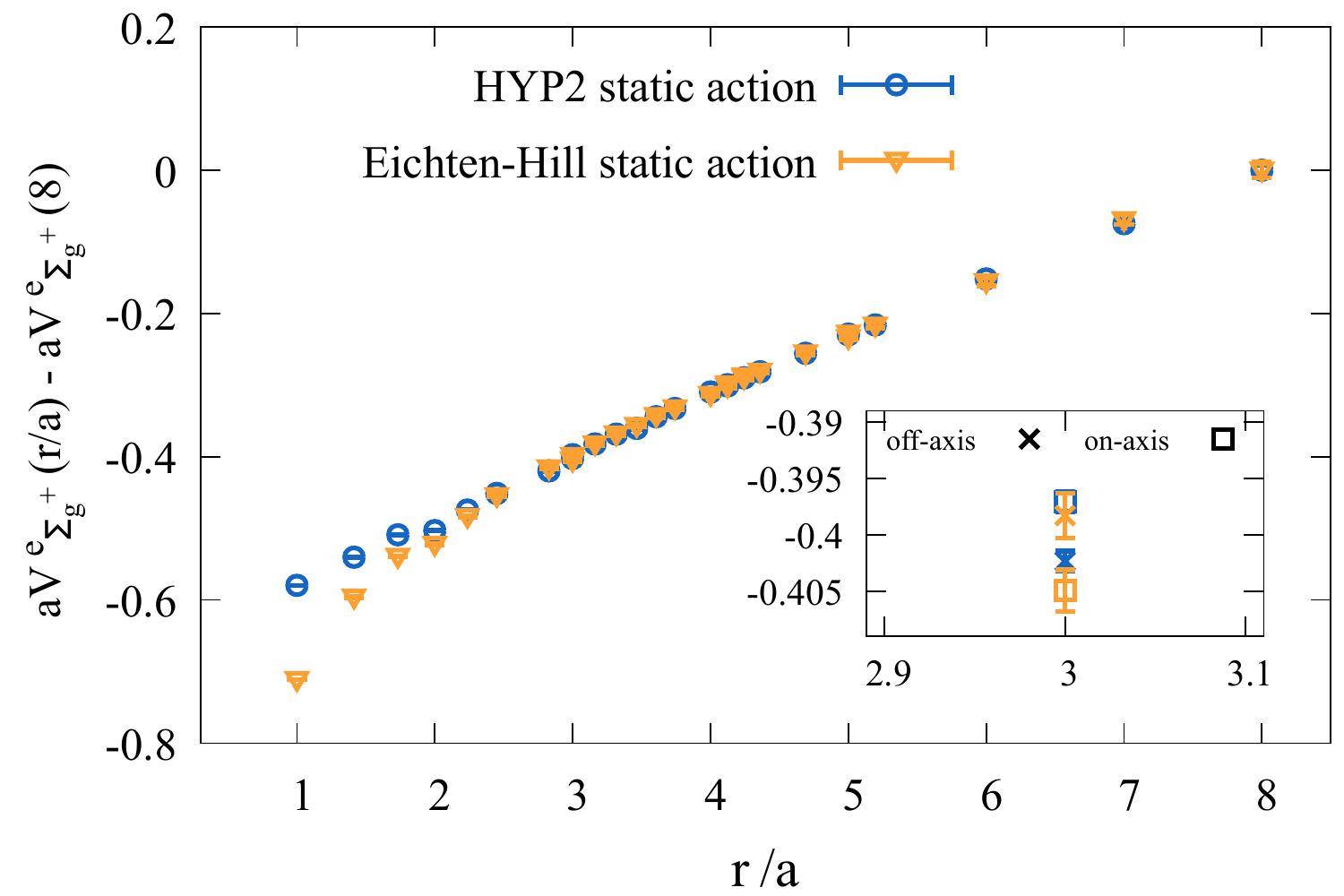}
			\end{minipage}
			\end{center}
		\begin{minipage}{0.5\linewidth}
			\subcaption{$r$-method: \\ $r \to r_\text{impr}$}\label{fig:rimproved}
			\centering\includegraphics[width=0.8\linewidth,page=2]{figure3}
		\end{minipage}
		\begin{minipage}{0.5\linewidth}
			\subcaption{$V$-method: \\ $V_{\Sigma_g^+}^e(r) \rightarrow V_{\Sigma_g^+}^e(r)-\Delta V_{\Sigma_g^+}^{\text{lat},e}(r)$}\label{fig:Vcorrected}
			\centering\includegraphics[width=0.8\linewidth,page=3]{figure3}
		\end{minipage}
		\caption{Unimproved and improved SU(2) lattice field theory data for the ordinary static potential at gauge coupling $\beta=2.40$ for two different discretizations of the static action.}
		\label{fig:treelevelimprov_methods}
	\end{figure}

	We now discuss and compare two commonly used methods (in the following referred to as $r$-method and $V$-method) to reduce lattice discretization errors for the static potential at tree level of perturbation theory.
	To assess the effectiveness of both methods, we plot lattice field theory data for the ordinary static potential for gauge group SU(2) at gauge coupling $\beta=2.40$ computed with two different discretizations of the static action,
	the HYP2 static action and the standard Eichten-Hill static action.
	The two discretizations should lead to similar results, where discrepancies are the consequence of lattice discretization errors.
	For the unimproved data shown in Figure~\ref{fig:unimproved}, discretization errors are rather large, particularly pronounced at small $r/a$. 
	These errors also cause a breaking of rotational symmetry, which is reflected by the discrepancy of data points from on-axis Wilson loops and from off-axis Wilson loops with the same spatial separation, e.g.\ $r/a=|\mathbf{r}|/a=3$ with $\mathbf{r}/a=(3,0,0)$ and $\mathbf{r}/a=(2,2,1)$.
	
	The $r$-method of improvement was introduced for the static force~\cite{Sommer:1993ce}.
  	The basic idea is to match the static force computed on the lattice at tree level of perturbation theory with the corresponding continuum result.
	Later, the $r$-method was also adopted for the static potential~\cite{Necco:2001xg}.
	A static potential data point is shifted from its original separation $r$ to an improved separation $r_\text{impr}$ defined via $({4 \pi r_\text{impr}})^{-1} = G(\textbf{r}/a)/a$,
	where $G(\textbf{r}/a)$ represents the lattice propagator at tree level, which depends on the discretization of the static action.
	The resulting improved data is presented in Figure~\ref{fig:rimproved}.
	There is still a sizable discrepancy resulting from an overcorrection of data points. 
	A universal parametrization of the two lattice data sets with small $\chi^2 / \text{dof} \approx 1$ is not possible.
	
	One can try to cope with the remaining discretization errors, e.g.\ by adding systematic errors to data points at small $r$ to reduce their weight in subsequent fits or by multiplying the data with a correction factor~\cite{Brambilla:2022het,Bazavov:2019qoo,Bazavov:2014soa,Komijani:2020kst}.
	We have explored a different strategy.
	First we have checked that the overcorrection can be consistently described by $\bar{\Delta}^\text{lat}=\sigma (r-r_\text{impr})$, where $\sigma$ denotes a fit parameter closely related to the string tension. 
	This expression can be motivated by noting that a Cornell ansatz $V_0 - \alpha/r + \sigma r$ is a reasonable description of the ordinary static potential and by assuming that one-gluon exchange is strongly related to the $\alpha/r$ term, but not to the other $r$-dependent term $\sigma r$.
	A significant part of the remaining discretization errors can be removed by subtracting the overcorrection term $\bar{\Delta}^\text{lat}=\sigma (r-r_\text{impr})$ from the $r$-improved lattice data points for the static potential.
	$\sigma$ should be determined by a fit of $V_0 - \alpha/r + \sigma r$ to lattice data points at larger $r/a$, where discretization errors are negligible.
	We plan to discuss this in more detail in an upcoming publication.
	We note that such an additional correction is not necessary for a tree-level improvement of the static force, since the problematic linear term $\sigma r$ in the static potential corresponds to a constant term in the static force which is independent of $r$.
	
	The alternative $V$-method~\cite{Michael:1992nj,Hasenfratz:2001tw} corrects the static potential by subtracting the difference between the lattice static potential at tree level, which is proportional to $G(\textbf{r}/a)/a$, and the continuum static potential at tree-level, which is proportional to $1/r$, from the full lattice static potential, i.e.\
	\begin{equation}\label{eq:improvedpotential}
		V_{\Sigma_g^+}^e(r) \rightarrow V_{\Sigma_g^+}^e(r)-\Delta V_{\Sigma_g^+}^{\text{lat},e}(r) = V_{\Sigma_g^+}^e(r) -  \alpha' \left(\frac{1}{r} - \frac{G^e(\textbf{r}/a)}{a}\right).
	\end{equation}
	$\alpha'$ is determined by a fit to the unimproved data as discussed in section~\ref{SEC005} and is related to the strong coupling constant.
	The improved data $V_{\Sigma_g^+}^e(r)-\Delta V_{\Sigma_g^+}^{\text{lat},e}(r)$ is presented in Figure~\ref{fig:Vcorrected}.
	It can be consistently described by a smooth curve and rotational symmetry is restored within statistical errors (see the zoomed plot in Figure~\ref{fig:Vcorrected}).

	The important conclusion of this section is that the $V$-method is clearly superior to the $r$-method when computing the static potential. On the contrary, for the static force we expect that both methods perform on a similar level.


	\section{\label{SEC005}Parametrizations of the hybrid static potentials}
\begin{figure}\centering
		\includegraphics[width=0.6\linewidth,page=2]{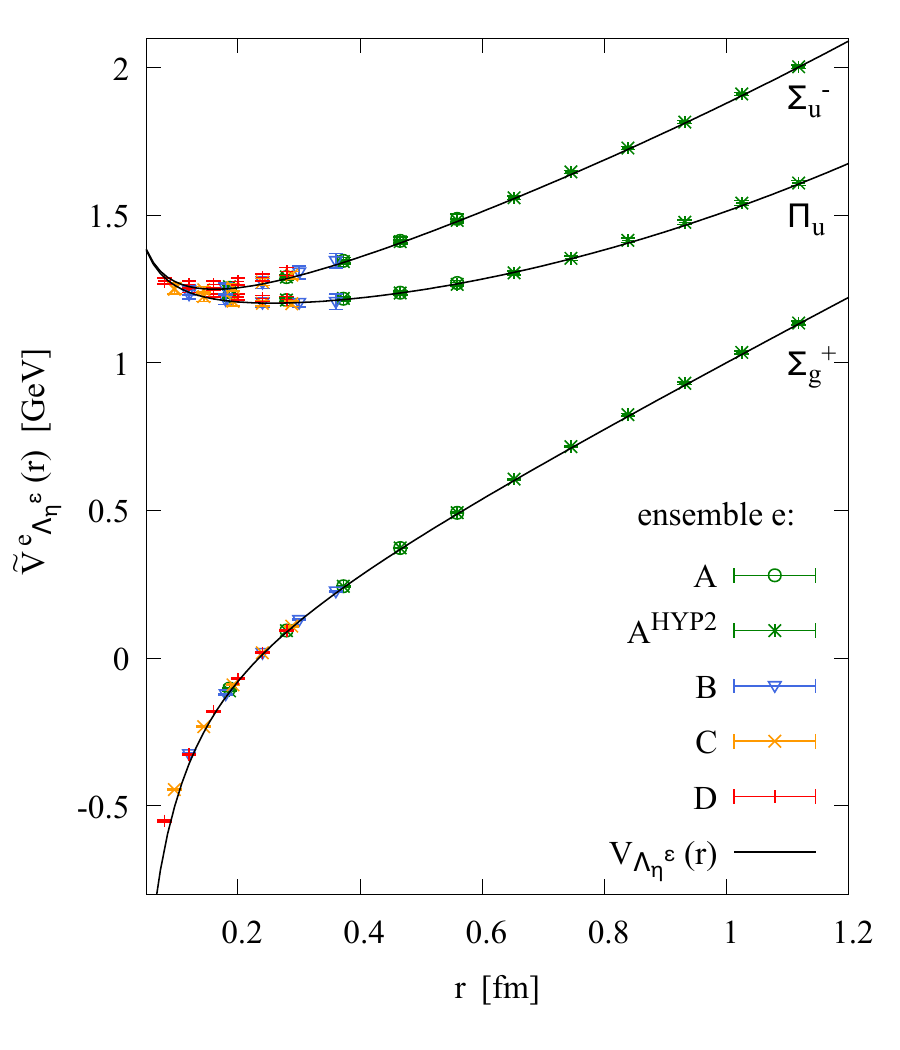}
		\caption{Improved lattice results for the ordinary static potential $\Sigma_g^+$ and the hybrid static potentials $\Pi_u$ and $\Sigma_u^-$ from the five ensembles $e \in \{A,\,B,\,C,\,D,\,A^\text{HYP2}\}$ and their parametrizations~ \eqref{eq:parametrization_Ordinary}, \eqref{eq:extendedparametrizationHybrid_Piu} and \eqref{eq:extendedparametrizationHybrid_Sigmau}.}
		\label{fig:improved_SU3data}
	\end{figure}

	To remove lattice discretization errors at tree level of perturbation theory in our SU(3) lattice data, we employ the $V$-method described in the previous section.
	We carry out a joint $8$-parameter fit to the lattice results for the ordinary static potential from all ensembles $e \in \{A,B,C,D,A^\text{HYP2}\}$.
	The fit function is
	\begin{equation}\label{eq:ordinary_potential}
		V_{\Sigma_g^+}^{\text{fit},e}(r) = {V_{\Sigma_g^+}(r)} + {C^e} +{\Delta V^{\text{lat},e}_{\Sigma_g^+}(r)},
	\end{equation}
	with the lattice discretization error at tree level, ${\Delta V^{\text{lat},e}_{\Sigma_g^+}(r)}$, as defined in Eq.~(\ref{eq:improvedpotential}) ($\alpha'$ is one of the fit parameters), the ensemble-dependent self energy of the static quarks, ${C^e}$, and the parametrization of the ordinary static potential,
	\begin{equation}\label{eq:parametrization_Ordinary}
		{V_{\Sigma_g^+}(r)} = -\frac{\alpha}{r} + \sigma r .
	\end{equation}
	Eq.~\eqref{eq:ordinary_potential} is fitted to the lattice data for $r \ge 0.2\,\text{fm}$.
	We define the improved lattice data points for the ordinary static potential via
	\begin{equation}\label{eq:def_V_Sigmagplus_tilde}
		\tilde{V}_{\Sigma_g^+}^e(r) = V_{\Sigma_g^+}^e(r) - C^e - \Delta V^{\text{lat},e}_{\Sigma_g^+}(r),
	\end{equation}
	where the self energy $C^e$ and the lattice discretization errors at tree level of perturbation theory are subtracted.
	This improved data is presented together with its parametrization~\eqref{eq:parametrization_Ordinary} in Figure~\ref{fig:improved_SU3data}.

	The lattice results for the hybrid static potentials $\Pi_u$ and $\Sigma_u^-$ from all ensembles can be described consistently by a $10$-parameter fit with
	\begin{equation}\label{eq:fit_hybrid_potential}
		V^{\text{fit},e}_{\Lambda_{\eta}^{\epsilon}}(r) =  {V_{\Lambda_{\eta}^{\epsilon}}(r)} + C^e +  {\Delta V^{\text{lat},e}_\text{hybrid} (r)} +  {A'^e_{2,\Lambda_{\eta}^{\epsilon}} a^2 }
	\end{equation}
	for $r\ge 2a$.
	For the hybrid static potentials the lattice discretization errors at tree level of perturbation theory are ${\Delta V^{\text{lat},e}_\text{hybrid} (r)} = -\frac{1}{8} \Delta V^{\text{lat},e}_{\Sigma_g^+}(r)$.
	The parametrizations ${V_{\Lambda_{\eta}^{\epsilon}}(r)}$ with $\Lambda_{\eta}^{\epsilon}=\Pi_u,\, \Sigma_u^-$ are based on a prediction of potential Non-Relativistic QCD for short quark-antiquark separations~\cite{Berwein:2015vca}.
	They are given by
	\begin{align}
		\label{eq:extendedparametrizationHybrid_Piu}
	 	{V_{\Pi_u}(r)} &= \frac{A_1}{r} + A_2 + A_3 r^2 \\
		\label{eq:extendedparametrizationHybrid_Sigmau}
		 {V_{\Sigma_u^-}(r)} &= \frac{A_1}{r} + A_2 + A_3 r^2 + \frac{B_1 r^2}{1 + B_2 r + B_3 r^2}
	\end{align}
	with fit parameters $A_1,\,A_2,\, A_3$, which are the same for both hybrid static potentials, and an additional term with fit parameters $B_1,B_2$ and $B_3$
	for the $\Sigma_u^-$ potential.
	The term $A'^e_{2,\Lambda_{\eta}^{\epsilon}} a^2$ in Eq.~\eqref{eq:fit_hybrid_potential} accounts for the discretization error of the constant shift with respect to the ordinary static potential at leading order in the lattice spacing.
	The fit parameter $A'^e_{2,\Lambda_{\eta}^{\epsilon}}$ with $\Lambda_{\eta}^{\epsilon}=\Pi_u$ or $\Sigma_u^-$ is equal for the ensembles $e \in \{A,B,C,D\}$ and different for the ensemble $e= A^\text{HYP2}$.
	
	As before, we define improved data points for the hybrid static potentials by subtracting the self energy $C^e$, the lattice discretization error at tree level of perturbation theory ${\Delta V^{\text{lat},e}_\text{hybrid} (r)}$ and the lattice discretization error of the constant shift $A_2$ at leading order in the lattice spacing, $A'^e_{2,\Lambda_{\eta}^{\epsilon}} a^2$, i.e.\
	\begin{equation}
		\tilde{V}^e_{\Lambda_\eta^\epsilon}(r) = V^e_{\Lambda_\eta^\epsilon}(r) - C^e - \Delta V^{\text{lat},e}_\text{hybrid} (r) - A'^e_{2,\Lambda_\eta^\epsilon} a^2 .
	\end{equation}
	The improved lattice data points and the corresponding parametrizations \eqref{eq:extendedparametrizationHybrid_Piu} are consistent within statistical errors (see Figure~\ref{fig:improved_SU3data}) and, thus, seem to represent continuum limit results for hybrid static potentials.

	Since we are considering several finer lattice spacings than before, we are able to reach short quark-antiquark separations as small as $0.08\,\text{fm}$.
	At these small separations the lattice results clearly exhibit the repulsive behavior predicted by perturbation theory and indicate the expected degeneracy for $\Pi_u$ and $\Sigma_u^-$.
	The parametrizations and improved lattice data points are provided in our related journal publication \cite{Schlosser:2021wnr} for straightforward use, e.g.\ for refined Born-Oppenheimer approaches to predict the spectra of heavy hybrid mesons.
	The parametrizations provided in this work are expected to change the mass spectra by $\order{10\dots 45\text{ MeV}}$ compared to parameterizations obtained previously at much coarser lattice spacing (for details see Ref.~\cite{Schlosser:2021wnr}).


\section*{Acknowledgments}
	We thank Christian Reisinger for providing his multilevel code.
	We acknowledge interesting and useful discussions with Eric Braaten, Nora Brambilla, Francesco Knechtli, Colin Morningstar, Lasse M\"uller, Christian Reisinger and Joan Soto.
	
	M.W.\ acknowledges support by the Heisenberg Programme of the Deutsche Forschungsgemeinschaft (DFG, German Research Foundation) -- project number 399217702.
	
	Calculations on the GOETHE-HLR and on the on the FUCHS-CSC high-performance computers of the Frankfurt University were conducted for this research. We thank HPC-Hessen, funded by the State Ministry of Higher Education, Research and the Arts, for programming advice.

\bibliographystyle{JHEP.bst}
\bibliography{LATTICE2022_hybridpotentials}

\end{document}